# Precision $\Upsilon$ and $J/\Psi$ spectroscopy with lattice NRQCD

NRQCD collaboration,
presented by C. T. H. Davies [a]*

[a]Department of Physics and Astronomy, University of Glasgow,
Glasgow, G12 8QQ, U.K.

We present new results on $\Upsilon$ and $J/\Psi$ spectroscopy using Lattice NRQCD. Charmonium results at quenched $\beta$ = 5.7 show agreement with experiment within understood systematic errors. We compare bottomonium results at quenched $\beta$s 6.0 and 5.7 and show that the $1P-1S$ splitting scales with $\Lambda_V$, provided a correction is applied for the $\mathcal{O}(a^2)$ errors inherent in the gluon field configurations. There is a clear scale difference between charmonium and bottomonium results in the quenched approximation. We compare quenched results at $\beta$=6.0 for bottomonium with those obtained on HEMCGC configurations at $\beta$=5.6 using 2 light flavours of dynamical quarks. We show that extrapolations to $n_f = 3$ are consistent with experiment for the ratio of the $2S - 1S$ splitting to that of the $1P - 1S$. We extract a value for the $\Upsilon$-$\eta_b$ splitting extrapolated to $n_f = 3$.

## 1. Introduction

The NRQCD collaboration has been active for a few years in the field of heavy quark physics. Our aim is to test QCD on the lattice by comparing charmonium and bottomonium spectroscopy to experiment as accurately as we can. This has required :

- The development of a non-relativistic effective action appropriate to the physics [1].

- A perturbative improvement scheme to remove discretisation errors order by order [2,3].

- High statistics calculations with multi-exponential fits to multiple smeared correlators [4–6].

Heavy-heavy spectroscopy lends itself very well to these techniques. An important feature is that sensible estimates can be made of the remaining systematic errors.

Previous results have focussed on $\Upsilon$ spectroscopy on quenched lattices at $\beta$=6.0 from the Staggered collaboration [6]. The new results reported below are on quenched lattices at $\beta$ = 5.7 (from UKQCD) and on lattices with 2 light flavours at $\beta$=5.6 (from HEMCGC). Viewing the NRQCD action as an expansion in powers of $v^2/c^2$ we have used, as before, leading and next-to-leading spin-independent terms along with leading spin-dependent terms. $v^2/c^2 \approx 0.1$ for $b\bar{b}$ and 0.3 for $c\bar{c}$. Discretisation errors in the leading order terms are also removed. The coefficients of all terms have been set to their tree-level values once the link fields have been 'tadpole-improved' by dividing by the fourth root of the plaquette [2,3].

## 2. Charmonium results

The charmonium spectrum as determined on quenched configurations at $\beta$=5.7 is shown in Figure 1 [7]. The bare lattice quark mass is fixed (at 0.8 in lattice units) from the requirement that the mass in the non-relativistic dispersion relation for the $\eta_c$ should be correct. The expected systematic error is 30 MeV from $v^6/c^6$ terms, and this error is clearly evident in the fine structure of the spectrum. Quenching corrections could also appear at this level in the hyperfine splitting.

The inverse lattice spacing, determined from the spin-averaged $1P - 1S$ splitting is 1.23(4) GeV. By spin-averaged splitting we mean $m(^1P_1) - 0.25 \times [3m(^3S_1) + m(^1S_0)]$. The $2S - 1S$ splitting gives a consistent value but is not well determined because excited states decay very quickly at low $\beta$.

---
*With other members of the NRQCD collaboration; K. Hornbostel (SMU), G.P.Lepage (Cornell), A.J.Lidsey (Glasgow), J.Shigemitsu (OSU), J.Sloan (SCRI).

## 3. Upsilon results

The bottomonium spectra that we obtain on quenched configurations at $\beta$=6.0 and $n_f$=2 configs at $\beta$=5.6 (HEMCGC) are shown in ref. [8]. The expected systematic error here is only 5 MeV. There are clear signs that the $2S-1S$ and $1P-1S$ splittings are inconsistent in the quenched approximation. This is a reflection of the scale-dependence of the inverse lattice spacing because the quenched coupling constant runs incorrectly. This problem is much improved on the $n_f$=2 configurations.

We can also compare the $1P - 1S$ splitting obtained at different values of $\beta$. Figure 2 shows this splitting divided by $\Lambda_V$ obtained from the plaquette [9] vs $\Lambda_V$ in lattice units. For asymptotic scaling the results at $\beta$ =6.0 and 5.7 should lie on a horizontal line. They do this, but only after a correction has been applied for the effect of $\mathcal{O}(a^2)$ effects from the plaquette gluon action. This correction can be calculated perturbatively [8] or non-perturbatively by comparing solutions in corrected and uncorrected heavy quark potentials [10]. This figure also makes clear that different $a^{-1}$ values are obtained from different quantities, even when discretisation errors are corrected. $a^{-1}(\Upsilon) > a^{-1}(\Psi) > a^{-1}(m_\rho)$.

Returning to a comparison of quenched and unquenched results, we show in Figure 2 the variation with $n_f$ of the ratio of the $2S - 1S$ splitting to that of the $1P - 1S$. All splittings are corrected perturbatively for the effect of gluonic $\mathcal{O}(a^2)$ errors. The corrections amount to only 2% at $\beta$=6.0. It is clear that the ratio is $n_f$ dependent. A linear extrapolation takes us to the experimental value of 1.28 at the value of $n_f$ appropriate to the momenta inside an $\Upsilon$ of 3. This is very gratifying and shows clearly that sensible values for physical quantities (e.g. $\alpha_s$ [8]) can be obtained by extrapolation in $n_f$. It would be useful to check the results for $n_f$=4.

In a similar way the hyperfine splitting $m(\Upsilon) - m(\eta_b)$ can be extrapolated to $n_f = 3$. We obtain a value of 44 MeV. The expected systematic error of 5 MeV dominates the error in this result. It may be true that $n_f$=4 is more appropriate to this short distance quantity and then one should correct it upwards perturbatively.

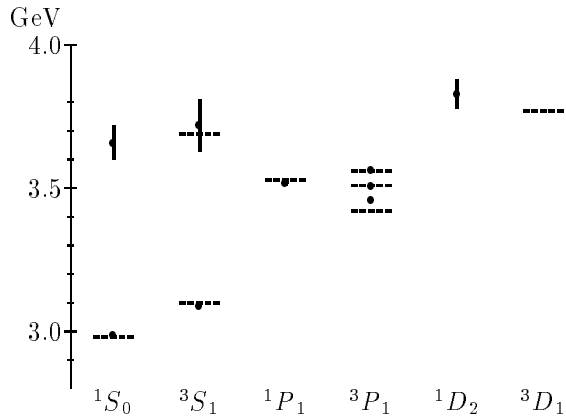

Figure 1. NRQCD simulation results for the charmonium spectrum plotted relative to the spin-average of the $J/\Psi$ and the $\eta_c(1S)$ using an inverse lattice spacing of 1.23 GeV from the $1P-1S$ splitting. Experimental values are indicated by dashed lines. Error bars are shown where visible, and only indicate statistical uncertainties.

## 4. Acknowledgements

We are grateful to the SCRI lattice gauge theory group for providing us with the HEMCGC configurations. We would also like to thank the Staggered collaboration for the use of their $\beta$=6.0 quenched configurations and the UKQCD collaboration for the use of their $\beta$=5.7 quenched configurations. This work was supported by DOE, NSF, PPARC and by the EU under contract CHRX-CT92-0051. The computing was done at OSC, NERSC and the Atlas Centre.

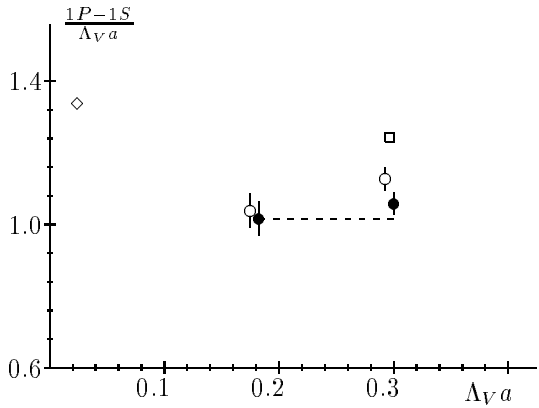
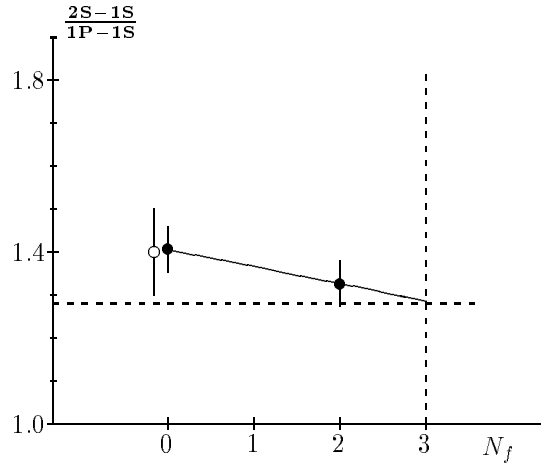

Figure 2. Asymptotic scaling of the $1P-1S$ splitting for $b\bar{b}$ for quenched results in the $V$ scheme. Open circles represent raw results at $\beta = 5.7$ and 6.0 and filled circles represent results corrected for gluonic $\mathcal{O}(a^2)$ errors. Filled circles are offset for clarity and the horizontal line is drawn to guide the eye. Two other points appear on the graph - the square from the $c\bar{c}$ spectrum at $\beta = 5.7$ and the diamond from the $a \to 0$ limit of $m_\rho$ from ref. 11. These two points have been rescaled by the ratio of the physical values of $1P - 1S$ for $c\bar{c}$ or $m_\rho$ to the $1P - 1S$ splitting of $b\bar{b}$ (taken to be 452 MeV).

Figure 3. Dependence on $N_f$ of the ratio of the $2S - 1S$ to $1P - 1S$ splittings in $b\bar{b}$. The filled circles give values at quenched $\beta = 6.0$ and $\beta=5.6$, $N_f=2$. The open circle is quenched $\beta =5.7$ for comparison. The horizontal dashed line gives the experimental value of 1.28 and solid line is an extrapolation to $N_f = 3$, drawn to guide the eye.